\documentstyle[11pt]{article}
\newcommand{\be}{\begin{equation}}
\newcommand{\ee}{\end{equation}}
\newcommand{\bq}{\begin{eqnarray}}
\newcommand{\eq}{\end{eqnarray}}
\newcommand{\n}{\noindent}

\begin{document}

\begin{titlepage}

\title{{\bf Some non perturbative calculations on spin~glasses}}
\author{Matteo Campellone}

\date{October, 1994}

\maketitle

\begin{center}
Dipartimento di Fisica, Universit\`a di Roma
  {\em La Sapienza}, \\
P. Aldo Moro 2,  00185 Roma,  Italy\\

\vspace{1truecm}
campellone@roma1.infn.it

\vspace{2truecm}

\begin{abstract}
Models of spin glasses are studied with a phase transition discontinuous in
the Parisi order
parameter. It is assumed that the
leading order corrections to the thermodynamic limit of the
high temperature free energy are due
 to the existence of a metastable saddle point in the replica
formalism.
 An ansatz is made on the
form of the metastable point and its contribution to the free
energy is calculated. The Random Energy Model
 is considered along with the $p$-spin and the
$p$-state Potts Models in their
$p < \infty$ expansion.
\end{abstract}

\end{center}

\vspace{4truecm}

Please contact me if you want the original pictures as poscript files.

\end{titlepage}

\vspace{3truecm}

\section{Introduction}
The mean field replica theory has revealed to be a very powerful method
 to study spin glass
models \cite{mpv}. The main problem with the replica method is
 that it necessarily requires an
ansatz on the solution. The Random Energy Model (R.E.M.)
 is a very simplified  spin glass model which
can also be solved without
 the utilisation of the replicas. It is, therefore, a very good
testing ground for
 the replica  method since, although very simple, it
presents the typical features of
 spin glasses such as replica symmetry breaking and
non-ergodicity. For $T$ greater than a
 critical temperature
$T_{c}$ the R.E.M. presents no breaking
 of the permutation symmetry among the replicas which, in
absence of external fields, tend to have
 the lowest possible value of the mean overlap.
When $T < T_{c}$ the system undergoes a phase
 transition towards a one-replica-symmetry-broken phase.
The corresponding latent heat is zero and so the transition is second order.
The R.E.M. was first introduced and solved by B.~Derrida who was also able to
calculate the finite-size
 corrections to the high temperature free energy \cite{rem1}.

In the present work
this last result will be obtained by making use of the replica method.
In doing so, a metastable saddle point
 in the replica space will be individuated as the responsible
for the leading order finite-size corrections.
The $p$-spin model and the $p$-state Potts model will also be considered.
Both these models tend to a R.E.M. in the limit $p \rightarrow
\infty$.

The scheme of this paper is as follows:
Section $1$ is based on reference \cite{rem1}; the Random Energy Model and
Derrida's results for the
 finite-size corrections will be presented. In section $2$ the same results
as the contribution
 of a metastable saddle point will be recovered. Section 3 will introduce
the
$p$-spin model and repeat
 the calculation performed for the R.E.M. in the formalism
of a
$(p=\infty)$-spin model. In section 4
 the results will be extended to the case $p < \infty$. In
section 5 the $p$-state Potts
 model will be introduced and the $p=\infty$ limit will be studied.
Finally, a $p < \infty$ expansion for
 the Potts model will be formulated and the calculations will
be extended to this case.

\section{The Random Energy Model}

It is worthwhile
recalling the main results on the R.E.M. in order to establish the notations.
The model describes the behaviour
 of any system with a fixed number of energy
levels whose energies are independently distributed according to a gaussian
law.
If $2^{N}$ is the number of levels, the model is defined by the properties:

\begin{equation}
P(E) = \frac{1}{\sqrt{N \pi J^2}}\exp{\left[\frac{-E^2}{NJ^2}\right]}
\label{distrprob}
\end{equation}

\begin{equation}
P(E_{i},E_{j}) = P(E_{i})P(E_{j}).
\end{equation}

\noindent
The solution can be easily derived using a microcanonical argument. For the
free
energy one finds:

\begin{equation}
F = \left\{\begin{array}{ll}
      N[-\frac{\ln{2}}{\beta} - \frac{\beta J^2}{4}] &\hbox{for $\beta < \beta
_{c}$ \ \ \ \ \ (a)}\\
     -E_{0}  &\hbox{for $\beta > \beta _{c}$ \ \ \ \ \ (b)},
    \end{array}
      \right.
\label{Frem}
\end{equation}

\n
where  $E_{0} = NJ \sqrt{\ln{2}}$ and $\beta _{c} = 2\sqrt{\ln{2}}/J$.

The R.E.M. can also be solved making use of the replicas.
If one computes the $n$-$th$ power of the partition function one obtains

\begin{equation}
Z^n = \sum_{\{p_{i}\}}\frac{n!}{\prod_{i}{(p_{i}!)}}
\exp{\left[-\sum_{i}^{2^N}{\frac{p_{i}E_{i}}{T}}\right]},
\label{znnonmediata}
 \end{equation}

\noindent
where

\begin{eqnarray}
p_{i}&\geq &0  \nonumber \\
\sum_{i}^{2^N}{p_{i}}& = &n.
\label{AA}
\end{eqnarray}

\noindent
After averaging over the disorder one has

\begin{equation}
\overline{Z^n}=\sum_{\{\nu\}}\frac{n!}{\prod_{p=1}^{n}{(p!)^{\nu_{p}!}}
\prod_{p=1}^{n}{\nu_{p}!}}
\exp{\left[N\sum_{p=0}^{n}{\nu_{p}\left[\ln{2} +
\frac{p^{2}\beta^2 J^2}{4}\right]}\right]},
\label{FF}
\end{equation}

 \noindent
where $\nu_{p}$ is the number of $p_{i}$ that are equal to $p$.
 The $\nu_{p}$ verify the conditions

\begin{eqnarray}
\nu_{p} &\geq & 0 \nonumber
\label{BB}  \\
\sum_{p=0}^{n}{\nu_{p}} &=& 2^N  \nonumber
\label{CC}   \\
\sum_{p=1}^{n}{p\nu_{p}} &=& n.
\label{DD}
\end{eqnarray}

\n
To obtain equation (\ref{FF}) it has
 been used the fact that for large enough $N$ it can be written

\begin{equation}
\frac{(2^N)!}{\nu_{0}!}\sim 2^{N\sum_{p=1}^{n}{\nu_{p}}}.
\end{equation}

\noindent
One can find that
 for $T > \sqrt{n}\,T_{c}$ the dominant contribution of (\ref{FF}) is obtained
by taking

\bq
&&\nu_{1}=n \\
&&\nu_{p\geq 2}=0.
\label{sstab}
\eq

\n
The correspondent expression for $\overline{Z^{n}}$ is linear in
$n$, so one can use the well known formula

\begin{equation}
\overline{\ln Z} =\lim_{n \rightarrow 0} \frac{\overline{Z^n}-1}{n}
\label{formularepliche}
\end{equation}

\noindent
to calculate the high
temperature free energy (\ref{Frem}(a)). The low temperature
expression of (\ref{Frem}) is instead obtained by taking
$\nu_{p\neq\bar{p}}=0$ and
 $\nu_{p=\bar{p}}=n/\bar{p}$, where $\bar{p}$ comes out to be equal to
$T/T_{c}$.

Finally one can write, without deriving it, the
expression for the finite size high temperature free energy.

\begin{eqnarray}
&\overline{\ln{Z}}& = N\left[\ln2 + \frac{J^2}{4T^2}\right] -
\frac{1}{2}\overline{\left[\frac{Z}{\overline{Z}} - 1\right]^2} +
\frac{1}{3}\overline{\left[\frac{Z}{\overline{Z}} - 1\right]^3} +
\cdots +
\frac{(-1)^{k+1}}{k}\overline{\left[\frac{Z}{\overline{Z}} - 1\right]^k}
 + \nonumber \\
&+&\frac{2T\sqrt{\pi}}
{J\left[\frac{T^2}{T_{c}^2} + 1\right]\sqrt{N}
\sin{\left(\frac{\pi}{2}\left[\frac{T^2}{T_{c}^2} + 1\right]\right)}}
\exp{\left[-\frac{NT^{2}J^{2}}{16}\left[\frac{1}{T_{c}^2} -
 \frac{1}{T^2}\right]^2\right]} +
\cdots  \nonumber  \\
\nonumber  \\
&&\hbox{per $\sqrt{2k-1} \,T_{c}<T< \sqrt{2k+1} \,T_{c}$}.
\label{Fcorr}
\end{eqnarray}

\n
Derrida derives expression (\ref{Fcorr})
 for $k =1,2,3$. He also makes the more general guess that it
should be true for every $k \geq 1$.

\section{The Metastable Point}
In this section the replica
 method shall be used to try to derive equation (\ref{Fcorr}).
It has been asserted that,
 for $T > T_{c}$, the dominant contribution to the equation (\ref{FF}) is
given by the choice
 (\ref{sstab}). It will now be considered the effect of only one grouping of
$m$ replicas. This can be done by taking

\begin{eqnarray}
\nu_{1}&=&n-m,  \nonumber\\
\nu_{m}&=&1.
\label{condasimm}
\end{eqnarray}

\noindent
For large $N$ equation (\ref{FF}) can thus be written as follows

\begin{equation}
\overline{Z^n} = \overline{Z^n}_{dom} + \overline{Z^n}_{sub} \ ,
\label{QQ}
\end{equation}

\noindent
where $dom$ stands for `dominant'
and $sub$ stands for `subdominant'. $\overline{Z^n}_{dom}$ is
given by the choice (\ref{sstab})
 in equation (\ref{FF}) while $\overline{Z^n}_{sub}$ is given
by the choice (\ref{condasimm}). One then has

\begin{equation}
\overline{\ln{Z}} = N\left[\ln2 + \frac{J^2}{4T^2}\right] +
\lim_{n \rightarrow0}\frac{1}{n} \overline{Z^n}_{sub},
\end{equation}

\noindent
where in $\overline{Z^n}_{sub}$ all the integer $m$ greater than $m=1$ are
 summed over.
One then has

\be
\lim_{n \rightarrow0}\frac{1}{n} \overline{Z^n}_{sub} = -
\sum_{m=2}^{\infty}\frac{(-1)^m}{m}\exp{N\left[(1 - m)\ln2 +
\frac{\beta^2 J^2}{4}(m^2 - m)\right]}.
\label{art1}
\end{equation}

\n
The sum in (\ref{art1}) can be written as an integral in the complex plane
over the circuit $({\cal C})$ as shown in figure

\n
One has

\begin{eqnarray}
\sum_{m=2}^{\infty}\frac{(-1)^m}{m}\exp{N\left[(1 - m)\ln2 +
\frac{\beta^2 J^2}{4}(m^2 - m)\right]} =  \nonumber \\
= -\frac{1}{2}\int_{{\cal C}}dm\frac{e^{N\left[(1-m)\ln2 +
\frac{\beta^2 J^2}{4}(m^2 - m)\right]}}{m\sin{(\pi m)}}.
\end{eqnarray}

\noindent
Both the sum and
 the integral are not well defined. They can be defined by deforming the
circuit
${\cal C}$ into a vertical path as indicated in figure 2. It can then be
written

\begin{eqnarray}
\sum_{m=2}^{\infty}\frac{(-1)^m}{m}\exp{N\left[(1 - m)\ln2 +
\frac{\beta^2 J^2}{4}(m^2 - m)\right]} \equiv
 -\frac{1}{2}\int_{\uparrow}dm\frac{e^{N\left[(1-m)\ln2 +
\frac{\beta^2 J^2}{4}(m^2 - m)\right]}}{m\sin{(\pi m)}}.
\end{eqnarray}

\n
The integral on the
 right hand side of the previous equation can be solved with the aid of the
saddle point method. Two remarks have to be made on this subject.
The first is that since the integration is
over imaginary $dm$, and, since
 the exponent is quadratic in $m$, the saddle point $m_{sp}$ is
given by the minimum and not by the maximum in $m$.
Therefore

\begin{equation}
m_{sp} = \frac{1}{2}\left(1 + \frac{T^2}{T_{c}^2}\right).
\label{msp}
\end{equation}

\n
The second remark
 is that to the saddle point contribution one has to add the residues of the
integrand that fall on
 the left of the $m_{sp}$ if the phase constant circuit $(\uparrow)$ shifts on
its right hand side when one makes it pass trough
$m_{sp}$.

\n
The evaluation of the integrand on
 $m_{sp}$ gives exactly the last term in equation (\ref{Fcorr}).
The terms $\frac{(-1)^{k+1}}{k}\overline{\left[\frac{Z}{\overline{Z}}
 - 1\right]^k}$ are reproduced
by the residues that are added.
 The number $k$ of those terms is given by the condition
$2 \leq k < m_{sp}$ which is equivalent to the condition $\sqrt{2k-1}
\,T_{c}<T< \sqrt{2k+1} \,T_{c}$ in equation (\ref{Fcorr}).
The ansatz
(\ref{condasimm}) seems
 therefore to be true, at least in the range of temperatures $T_{c} < T <
\sqrt{7} \,T_{c}$, where expression (\ref{Fcorr}) has been proved
correct by Derrida.
Thus, the choice (\ref{condasimm})
 (\ref{QQ}) reproduces exactly expression (\ref{Fcorr}) while other
possible choices of the $\nu_{p}$
 seem to lead to lower order contributions. In reference
\cite{tesi} this assertion
 could be proved rigorously for choices of the $\nu_{p}$ such as

\begin{eqnarray}
\nu_{1}&=&n - m -r, \nonumber \\
\nu_{m}&=&1  \hspace{.3 in}\mbox{with $m > 0$}, \nonumber \\
\nu_{r}&=&1 \hspace{.3 in} \mbox{with $r > m$}.
\label{condasimmbis}
\end{eqnarray}

\n
It can also be argued
 that this is likely to happen in general for more complicated choices of
the
$\nu_{p}$ where
 the insertion of more groupings is allowed. Furthermore, these arguments do
not
seem to be dependent on $k$.
 If this is so, the choice (\ref{condasimm}) provides the leading order
corrections to the high
 temperature free energy for all $T > T_{c}$ and expression (\ref{Fcorr}) is
correct even for $k>3$.

\section{The $p$-Spin Model}
All this may now be
 extended to the $p$-spin model which is defined by the Hamiltonian

\be
[A{\cal H}_{p}(\{s\}) =
 -\sum_{(1\leq i_{1}<i_{2}<\cdots<i_{p}\leq N)}\!\!\!\!\!\!\!\!\!\!\!\!\!
J_{i_{1},i_{2},\cdots,i_{p}} s_{1}\cdots s_{p}  +  h\sum_{i}s_{i},
\label{Hpspin}
\ee

\noindent
where $h$ is an external
 magnetic field and the $J_{i_{1},i_{2},\cdots,i_{p}}$ are random variables
that obey the gaussian law

\be
P(J_{i_{1},i_{2},\cdots,i_{p}}) = \left[\frac{N^{p-1}}{\pi J^2
p!}\right]^{-\frac{1}{2}} \exp{\left[-\frac{(J_{i_{1},i_{2},\cdots,i_{p}})^2
N^{p-1}}{J^2 p!}\right]}.
\label{distrprobp}
\ee

\n
For sake of simplicity, from now on it will be assumed $J=1$.
If one indicates with $E_{i}$ the
 energy relative to configuration $\{\sigma \}_{i}$, it is a well
known result that

\be
P(E_{1},E_{2},\cdots,E_{k}) \stackrel{p\rightarrow\infty}{\longrightarrow}
\prod_{i=1}^{k}P(E_{i}) \hspace{.3 in} \mbox{for $|q^{i,j}|<1$,\,$\forall
i,j$},
\label{fattorizz}
\ee

\n
where $q_{i,j}$ indicates the
 overlap between $(\{\sigma\}_{i})$ and $(\{\sigma\}_{j})$. In the
limit $p \rightarrow \infty$ then, the $p$-spin model reduces to R.E.M..

The expression for $\overline{Z^{n}}$ is

\be
\overline{Z^{n}} = \sum_{\{s_{i}^{a}\}}
\exp{
     \left[
     \frac{\beta^{2} N}{4}
     \left(
           n + \sum_{a \neq b}{Q_{ab}^{p}(s)}
      \right)
      \right]
     },
\ee
\noindent
where $Q_{ab}(s) \equiv
\frac{1}{N}\sum_{i}^{N}{s_{i}^{a}s_{i}^{b}}$.

Above $T_{c} = 1/2\sqrt{\ln{2}}$
 the model presents no breaking of the replica symmetry. After the
elimination of all the
 requested Lagrange multipliers, the high temperature solution is a $n
\times n$ matrix of the form

\begin{eqnarray}
Q_{ab} &=& q_{0}, \hspace{.3in} \mbox{for $a\neq b$},  \nonumber   \\
Q_{aa} &=&  0,
\end{eqnarray}

\n
For $T < T_{c}$ the matrix
 $Q_{ab}$ has two parameters $q_{0}$ and $q_{1}$. In the limit $p
\rightarrow \infty$ the solution
 is $q_{0} = 0$ and $q_{1} = 1$ and one recovers expression
(\ref{Frem}).
 For simplicity it has been put $h=0$ because it will not affect the point.

Assuming now $T > T_{c}$, one
 can proceed equivalently to what has been done in the previous section
and evaluate
$\overline{Z^{n}}$ on a one block matrix
$Q_{ab}$ of the form

\begin{equation}
Q_{1Block}=\left(
\begin{array}{c c c c c c c c c c }

0 & q_1 & q_1& q_1 &q_0&q_0 &q_0 &q_0  \\
q_1 & 0 & q_1 & q_1&q_0&q_0 & q_0& q_0  \\
q_1 & q_1& 0 & q_{1} &q_0&q_0 & q_0  & q_0 \\
q_1 & q_1& q_1 &0& q_0 & q_0& q_0&q_0  \\
q_0 & q_0& q_0&q_0&0&q_0& q_0& q_0   \\
q_0 & q_0& q_0&q_0&q_0& 0& q_0& q_0  \\
q_0 & q_0& q_0&q_0&q_0& q_0& 0& q_0  \\
q_0 & q_0& q_0&q_0&q_0& q_0&q_0& 0
\end{array}
\right)
\label{q1blocco}
\end{equation}

\n
$Q_{1Block}$ is a $n \times n$
 matrix with an $m \times m$ block of elements $q_{1}$. To
represent $Q_{1Block}$
 it has been chosen $n = 8$ and $m=4$. It is easy to see that

\be
\lim_{n\rightarrow 0}\overline{Z^n}_{1Block} =
 \lim_{n\rightarrow 0}\overline{Z^n}_{sub}
= \lim_{n\rightarrow 0}
\sum_{m=2}^{\infty} \frac{n!}{(n-m)!m!}e^{-N\left[(1 - m)\ln{2} +
\frac{\beta^2}{4}(m^2 - m)\right]},
\ee

\n
which is the same
 result found in the previous section. Ansatz (\ref{condasimm}) is
therefore equivalent
 to a matrix of the form (\ref{q1blocco}). It is worth remarking that to obtain
$\overline{Z^n}_{1Block}$ one
 has to sum over all the possible ways of inserting an
$m \times m$ block in an $n \times n$ matrix.

This result will now be extended to the case $p < \infty$ in the fashion of
\cite{pspin}.
Before that, it
 is appropriate to brief\hspace{1pt}ly show the low temperature behaviour of
the
model. Under
 $T_{c}$, the $p < \infty$ expansion leads to the mean field equations

\be
q_{0} = 0, \hspace{.3 in}
 q_{1} = 1 - \frac{m \xi(m)}{(1-m)}\frac{e^{-\frac{p \beta^{2}m^2}{4}}}{2
\sqrt{p \beta^2 /2}},
\label{q0q1pert}
\ee

\n
where it has been set

\be
\xi(m) \equiv
 \frac{1}{\sqrt{2\pi}}\int_{-\infty}^{\infty}\!dz\,(2\cosh{(mz)} - 2^m
\cosh^{m}(z)).
\label{defxi}
\ee

\n
One also has

\be
T_{c} = \frac{1}{2\sqrt{\ln{2}}}\left(1
+ 2^{-(p+1)}\sqrt{\frac{\pi}{4p(\ln{2})^{3}}}\right).
\label{EEE}
\ee

\n
Equation (\ref{EEE}) for the critical
 temperature comes from the condition that the value of the
break point in the order parameter function
 \cite{p80b} that maximises the free energy is 1. This is
because of the nature of
 the transition which, though second order in the thermodynamic sense, is
discontinuous from the order parameter point of view \cite{pspin}.

Assuming now to set $T > T_{c}$,
 one can calculate $\overline{Z^n}_{sub}$ with the insertion of a
block as it was done in the $p \rightarrow \infty$ limit.
Defining $T_{c}^{\infty}~\equiv~1/(2\!\sqrt{\ln\!2})$, the finite-$p$
equivalent of expression (\ref{Fcorr}) can be easily obtained

\bq
&\overline{\ln{Z}}& = N\left[\ln2 + \frac{1}{4T^2}\right] -
\frac{1}{2}\overline{\left[\frac{Z}{\overline{Z}} - 1\right]^2} +
\frac{1}{3}\overline{\left[\frac{Z}{\overline{Z}} - 1\right]^3} +
\cdots +
\frac{(-1)^{k+1}}{k}\overline{\left[\frac{Z}{\overline{Z}} - 1\right]^k}
 + \nonumber \\
&+&\frac{2T\sqrt{\pi}
\exp{\left[-\frac{NT^{2}}{16}\left[\frac{1}{(T_{c}^{\infty})^2} -
\frac{1}{T^2}\right]^2 - N \eta + \frac{N \omega^2
\beta^2}{4}\right]}} {\left[\frac{T^2}{(T_{c}^{\infty})^2} + 1 +
2 \omega \right]\sqrt{N}
\sin{\left(\frac{\pi}{2}\left[\frac{T^2}{(T_{c}^{\infty})^2} + 1 +
2\omega\right]\right)}} +
\cdots  \nonumber  \\ \nonumber
 \\ &&\hbox{for $\sqrt{2k^{\prime}-1} \,(T_{c}^{\infty})<T<
\sqrt{2k^{\prime}+1} \,(T_{c}^{\infty})$},
\label{enliberacorr1}
\end{eqnarray}
\noindent

where

\be
k^{\prime} \equiv k - w .
\ee

\be
\omega \equiv 2 T^{2} \eta^{\prime}(m),
\label{defomega}
\ee

\be
\eta (m) \equiv
  \frac{\xi(m) e^{\frac{\beta^{2} p q_{i}^{p-1}m^2}{4}}}{\sqrt{2 \beta^{2} p
q_{i}^{p-1}}}.
\label{III}
\ee

\n
One also has

\begin{equation}
m_{sp} = \frac{1}{2}
 \left( 1 + \left(\frac{T}{T_{c}^{\infty}} \right)^2\right) +
\omega,
\label{mcorr}
\end{equation}
It is worthwhile noting that,
assuming $m_{sp} = 1$ in equation (\ref{mcorr}) one obtains the
equation (\ref{EEE}) for the critical temperature.
This means that one
 can find $T_{c}$ as the temperature at which the block inserted
in the metastable
 matrix disappears and $Q_{1Block}$ coincides
 with the stable saddle point matrix
of $\overline{Z^n}_{dom}$.
 Furthermore, equation (\ref{q0q1pert}) for $q_{1}$ is recovered
as a saddle point
 equation for $\overline{Z^n}_{sub}$. In absence of magnetic field, it has
been assumed
$q_{0} = 0$.

%*************%

\section{The Potts Model}

The results obtained up to
 now will be extended in this section to the Potts model which is defined
by the Hamiltonian

\be
{\cal H} = -\frac{1}{2}\sum_{i,j \neq 1}^{N}J_{ij}(p \delta_{p(i)p(j)} - 1)
- h_{\lambda} \sum_{i =
1}^{N}(p
\delta_{p(i),\lambda} - 1),
\label{hpotts}
\ee

\n
where $p(i) = 0,1,\dots,p-1$,
and the $J_{ij}$ are as usual random variables obeying a gaussian
distribution with variance
$1/N$ and mean value $J_{0}/N$.

For every $p > 4$, the
 model undergoes a phase transition of the same kind as the one observed in
R.E.M., from a replica
 symmetric phase to a one-symmetry-broken phase. In this work all the
ferromagnetic order parameters
will be neglected and the focus will be put only on the glassy
aspects of the model.
\n
Hence, the free energy is \cite{potts1},\cite{tesi}

\be
\frac{n \beta F(Q)}{N}
= -n \frac{\beta^2}{4}(p-1) + \frac{\beta^2}{2 p^2} \sum_{\alpha <
\beta}^{n}
\sum_{r,s}^{p}(q_{rs}^{\alpha \beta})^2 - \ln Z(Q),
\label{4D}
\ee

where

\be
\ln Z(Q) = \ln \left \{ \sum_{ \{ p(\alpha) \} } \exp
\left[\beta^2 \sum_{\alpha <
\beta}^{n} q_{p(\alpha)p(\beta)}^{\alpha \beta} \right] \right \},
\ee

and
\be
q_{rs}^{\alpha \beta} =
 q^{\alpha \beta}(p \delta_{r,s} -1), \ \ \ \ \ \ \ \ \ \ \ \mbox{with  $0
\leq |q^{\alpha \beta}|
\leq 1$}.
\ee

\n
According to the
 conventional wisdom, one can assume that in the high temperature phase one
has $q_{0} = 0$. Therefore the free energy is

\be
F = - N \left[ \frac{\beta}{4}(p-1) + \frac{\ln{p}}{\beta} \right].
\ee
and the entropy
\be
S = N \left[ \ln{p} - \frac{p-1}{4T^2} \right],
\ee

\n
which becomes negative if
 $T < T_{c} \equiv \sqrt{\frac{p-1}{4\ln{p}}}$. Under $T_{c}$ the
solution is given by
 a one-symmetry-broken matrix of elements $q^{\alpha \beta}$.

To calculate the low
 temperature free energy it is useful to use the $p$ vectors $\vec{e^{a}}\
(a=1,\cdots,p)$, defined by the relations

\be
e_{i}^a e_{i}^b = p \delta_{a,b} -1 \ \ \ \ \mbox{$i= 1,\cdots,p-1$}
\label{4B},
\ee

\n
where repeated indexes are summed.
If $q_{0}=0$ e $q_{1} = q$,
 with a little algebra, one gets to the low temperature free energy
expression

\bq
&& \frac{\beta F}{N} =
 - \frac{\beta^2}{4}(p-1) + \frac{\beta^2}{4}(p-1)(m-1)q^2  +
+  \frac{\beta^2}{2} (p-1) q+ \nonumber \\
&&-\frac{1}{m} \ln \int \frac{d\vec{z} \,
 e^{(-z^2/2)}}{\sqrt{2 \pi}} \left( \sum_{b =1}^{p}
\exp
\left[ k \, \vec{e^b} \cdot
\vec{z}\right] \right)^m,
\eq

\n
where $\vec{z} = (z_{1},z_{2},\cdots,z_{p-1})$ and $k \equiv \beta \sqrt{q}$.

In the limit $ p \rightarrow \infty$,
 the last integral can be exactly solved \cite{tesi} and the
solution is found to be given by the mean field equations

\begin{eqnarray}
&q_{0}& = 0  \ \ \ q_{1} =1 \nonumber \\
&T_{c}& = \frac{1}{2}\sqrt{\frac{p-1}{\ln{p}}}. \nonumber \\
&F & = \left\{\begin{array}{ll}
      N[-\frac{\beta}{4}(p-1) - \frac{\ln{p}}{\beta}]
 &\hbox{for $T > T_{c}$ \ \ \ \ \ (a)}\\
     -E_{0} = - N \sqrt{(p-1) \ln{p}} &\hbox{for $T < T_{c}$ \ \ \ \ \ (b)}
    \end{array}
      \right.
\label{4L}
\end{eqnarray}

\n
This solution describes a $p^{N}$ states
 Random Energy Model with variance proportional to
$(p-1)$. The finite size corrections
 for this limit are already known, so the more general case in
which the finite-$p$ corrections are included shall be treated directly.

\n
One can define

\be
\frac{\xi(m)}{p \beta \sqrt{q}} \equiv \int_{-\infty}^{\infty} \prod_{b=1}^{p}
da_{b}e^{-\frac{p}{2}(\sum_{b}a_{b}^2)}\delta
\left(\sum_{b=1}^{p}a_{b}\right)
 \left [ \left(\sum_{b}^{p}e^{\beta \sqrt{q}pma_{b}}\right) -
\left(\sum_{b}^{p}e^{\beta \sqrt{q}pa_{b}}\right)^{m} \right ],
\ee

\n
where obviously one has

\be
\xi(1) = 0.
\label{4O}
\ee

\n
Recent work has been done on the
 numerical estimation of the $p$-dimensional integral $\xi(m)$
\cite{Felix}. Correcting
 the integral in the free energy expression one obtains the mean field
equations

\be
q = 1 - \frac{1}{(\beta^3 p^2 (p-1)\sqrt{q}} \frac{\xi(m)}{m(1-m)} \left[
 m^2 \beta^2 (p-1) + \frac{1}{q} \right] e^{-\frac{\beta^2}{2} m^2 (p-1)q}.
\ee

\n
Assuming $\ln{p} >> 1$ the last term in square brackets can be neglected

\be
q = 1 - \frac{\xi(m)}{(1 - m)}\frac{\beta_{c}}{p^4 \beta^{2}} .
\label{4P}
\ee
\n
Setting $q = 1-\epsilon$, one finds an
 equation for the critical temperature

\be
\frac{1}{T_{c}^2} =
 \frac{4}{p-1} \left(\ln{p} + \frac{T_{c} \xi^{\prime}(1)}{p^4} \right)
+ O(\varepsilon^2),
\ee

\n
The finite size corrections are obtained
 proceeding in the same way as for the $p$-spin.
Defining

\bq
\eta(m)&\equiv& \frac{\xi(m)}{\beta (p-1) p^2} e^{-\frac{\beta^2}{2} m^2 (p-1)}
\nonumber
\\
\omega&\equiv& 2 T^{2} \eta^{\prime}(m),
\eq

\n
one gets

\begin{equation}
m_{sp} =
 \frac{1}{2} \left( 1 + \left(\frac{T}{T_{c}^{\infty}} \right)^2\right) +
\omega,
\end{equation}

\n
where $T_{c}^{\infty} \equiv \frac{1}{2} \sqrt{\frac{p-1}{\ln{p}}}$.

\n
and finally

\bq
&\overline{\ln{Z}}& = N\left[\ln2 + \frac{1}{4T^2}\right] -
\frac{1}{2}\overline{\left[\frac{Z}{\overline{Z}} - 1\right]^2} +
\frac{1}{3}\overline{\left[\frac{Z}{\overline{Z}} - 1\right]^3} +
\cdots +
\frac{(-1)^{k+1}}{k}\overline{\left[\frac{Z}{\overline{Z}} - 1\right]^k} +
\nonumber \\
&+&\frac{2T\sqrt{\pi}
\exp{\left[-\frac{NT^{2}}{16}\left[\frac{1}{(T_{c}^{\infty})^2} -
\frac{1}{T^2}\right]^2 - N \eta + \frac{N \omega^2
\beta^2}{4}\right]}}
{\left[\frac{T^2}{(T_{c}^{\infty})^2} + 1 + 2 \omega \right]\sqrt{N}
\sin{\left(\frac{\pi}{2}\left[\frac{T^2}{(T_{c}^{\infty})^2} + 1 +
2\omega\right]\right)}} +
\cdots  \nonumber
  \\ \nonumber  \\ &&\hbox{for $\sqrt{2k^{\prime}-1} \,(T_{c}^{\infty})<T<
\sqrt{2k^{\prime}+1} \,(T_{c}^{\infty})$},
\label{enliberacorr2}
\end{eqnarray}
\noindent
where
\be
k^{\prime} \equiv k - w .
\ee

\n
In principle, all this
 is equivalent to what has been done for the $p$-spin and all the
considerations made at the end of section 4 can be repeated here.

\section{Conclusion}

For the high temperature phase of the Random Energy Model, the partition
function was evaluated on a metastable point that
 was introduced in order to account for the
probability that a group of
$m$ replicas freezes in a phase
 which resembles the low temperature one. In this way the finite size
corrections to the free
 energy were calculated. The result was checked with the one obtained by
B.~Derrida without the use of replicas.
 The two approaches are totally independent.
Derrida proves his result to be true only in the range of temperatures
$T_{c}< T <\sqrt{7}\, T_{c}$.
 The results obtained in this work coincide with Derrida's, in
particular they coincide for
$T_{c}< T < \sqrt{7}\, T_{c}$.
 The reliability of this method does not seem to depend on the
temperature, provided that $T > T_{c}$.
Therefore, the equivalence of the two results in the range
where formula (\ref{Fcorr})
 can be proved to be true, seems to indicate the reliability of the result
for all temperatures. In extending this ansatz to the
$p$-spin and Potts models
 it was possible to identify a one-block matrix as the metastable point.
Mean field equations give,
 for the elements of the block, the same value as the low temperature
mean overlap. Furthermore, a
$p < \infty$ expansion was performed
 for these models in order to extend the results to finite-$p$
$p$-spin and Potts models.

\par\noindent

\vskip1truecm
\par\noindent
{\Large \bf Acknowledgments}
\par\noindent

\n
I am extremely grateful to Giorgio Parisi for his constant support.
I also would like to thank Enzo Marinari for his comments and suggestions.

% BIBLIOGRAFIA

% BIBLIOGRAFIA
\bibliographystyle{IEEE}

\end{document}